# Quantized Exciton–Exciton Annihilation in Monolayer $WS_2$ on $SrTiO_3$ Substrates with Atomically Flat Terraces

Yuto Kajino,[1,*] Kohei Sakanashi,[2] Nobuyuki Aoki,[2,3] Kenji Watanabe,[4] Takashi Taniguchi,[5] Kenichi Oto,[1,3] and Yasuhiro Yamada[1,†]

[1] Department of Physics, Chiba University, Inage, Chiba 263-8522, Japan

[2] Department of Materials Science, Chiba University, Inage, Chiba 263-8522, Japan

[3] Molecular Chirality Research Center, Chiba University, Chiba, 263-8522, Japan

[4] Research Center for Functional Materials, National Institute for Materials Science, 1-1 Namiki, Tsukuba 305-0044, Japan

[5] International Center for Materials Nanoarchitectonics, National Institute for Materials Science, 1-1 Namiki, Tsukuba 305-0044, Japan

Corresponding authors: [*]kajino@chiba-u.jp and [†]yasuyamada@chiba-u.jp

Abstract

Monolayer materials are strongly affected by their potential fluctuations, which are induced by an intrinsic corrugation or the surface roughness of the substrate. We compare the effective exciton-exciton annihilation (EEA) rate constants of monolayer $WS_2$ on substrates with different surface topographies. We show that monolayer $WS_2$ on the substrate with atomically flat terraces displays small effective EEA rate constant deviating from the overall tendency and multiple exciton decay components, which cannot be accounted for by a conventional EEA model. To obtain a correct description, we use a quantized EEA model. The intrinsic EEA rate constant for the flat-terrace substrates determined by this new model is comparable to that of hBN-encapsulated monolayer $WS_2$.

Atomically thin layers of materials, such as graphene or transition metal dichalcogenides (TMDCs), have been extensively studied owing to their intriguing optoelectronic properties and spin-valley phenomena [1–4]. Furthermore, the highly flexible nature of monolayers and multilayers enable us to fabricate them on various materials and develop the artificial materials by mechanical stacking of different types of layers, known as van der Waals (vdW) heterostructures [5,6]. On the other hand, monolayer materials are sensitive to the surrounding material, which offers external control over their physical properties by the dielectric screening [7–11], strain tuning [12,13], and magnetic exchange field [14,15]. The surrounding environment of a monolayer material can also give rise to spatial fluctuations of the electronic potential through spontaneous or artificial corrugation, which disturbs the carrier transport and causes an inhomogeneous broadening of optical transitions [16–19]. Therefore, methods have also been developed to reduce such fluctuations. For instance, it has been shown that hexagonal boron nitride (hBN)-encapsulation can reduce the corrugation of monolayer materials and can help to realize a spatially uniform electronic potential, which significantly improves the electronic and optical characteristics of the monolayer materials [20–24]. Recent experimental studies on Moiré superstructures verified the significance of the spatial modulation of the electronic potential for the control of properties in such layers and their heterostructures [25,26].

Potential fluctuations also play an important role in the intrinsic recombination processes of excitons. It has been reported that the exciton-exciton annihilation (EEA) rate constant of hBN-encapsulated monolayer $WS_2$ is about two orders of magnitude lower than that of monolayer $WS_2$ on commonly used $SiO_2$/Si substrates [27,28]. Such a discrepancy of the EEA rate constant has been explained by a model that considers a strong spatial variation of the exciton density; the effective EEA rate constant is enhanced by a local enhancement of the exciton density due to the substrate-induced non-uniform electronic potential [27–30]. This suggests that the optical processes in monolayer TMDCs can be controlled by modulating the surface corrugation, which would be beneficial for the design of functional TMDC-based heterostructure. Therefore, it is essential to develop a model that can describe how the exciton relaxation dynamics are altered by potential fluctuations and verify it experimentally.

In this Letter, we investigate the intrinsic exciton recombination dynamics in monolayer $WS_2$ fabricated on various substrates by time-resolved photoluminescence (PL) spectroscopy at room temperature. The effective EEA rate constant obtained by employing the conventional EEA model strongly depends on the surface topography of the substrates. In particular, monolayer $WS_2$ fabricated on $SrTiO_3$ substrate with atomically flat terraces has the effective EEA rate constant which is slightly lower than that of the hBN-encapsulated sample, and its PL dynamics have multiple decay components whose lifetimes are quantized. Here, we propose to employ a quantized EEA (QEEA) model and show that it can describe the experimental results well. With this model, we are able to estimate the size of the spatial fluctuation of the electronic potential quantitatively. Furthermore, we find that the intrinsic EEA rate constant estimated from the QEEA model is close to that of hBN-encapsulated monolayer $WS_2$, which has a spatially uniform potential [27]. Our experiments and analyses highlight the essential role of the potential modulation in nonlinear exciton recombination dynamics in monolayer materials.

The monolayer $WS_2$ was fabricated on five kinds of substrate by mechanical exfoliation from a bulk crystal. The substrates are, in order from low to high surface roughness, a $SrTiO_3$ step-and-terrace (ST), an $Al_2O_3$ ST, a normal $SiO_2$/Si, a normal $Al_2O_3$, and a normal $SrTiO_3$ substrate (see Supplemental Material). The number of layers was identified by PL, reflectance, optical microscopy, and atomic force microscopy (AFM) [11]. The substrates were purchased and used as received. The ST substrates have terraces with a typical width of about 100 nm and step heights equal to the size of the unit cell of the substrate material. The ST and normal substrates are different only in terms of surface roughness, which was examined by AFM (see Supplemental Material). The hBN-encapsulated monolayer $WS_2$ was prepared by the all-dry stamping method, which is described elsewhere [31]. The samples were not annealed after fabrication except for the hBN-encapsulated sample. Because the surface of the normal $SiO_2$/Si substrate exhibits a maximum roughness depth of about 1 nm and a mean peak separation of about 10 nm, the interface between the substrate and the monolayer affects the physical properties of the monolayer $WS_2$ through the potential fluctuations induced by local bending [17,32]. The ST substrates should induce a smaller potential modulation because of the atomically flat terraces. Here, the weak and uniform potential modulation caused by an intrinsic corrugation of the monolayer or the periodic step of the substrate is expected to be dominant.

Micro-PL spectroscopy was performed at room temperature. As excitation light source, we used light at 532 nm from a supercontinuum light source with a repetition rate of 10 MHz and a pulse width of a few picoseconds. By using a 50x objective lens, we achieved an excitation spot with a diameter of approximately 1 μm on the sample. The PL decay dynamics were obtained by a time-correlated single-photon counting technique with a time resolution of ~10 ps, which is sufficiently shorter than the observed PL decay time constants.

To examine the substrate dependence of the exciton recombination dynamics, we first estimated the effective EEA rate constants from the experimental results by using the conventional EEA model where a homogeneous exciton density is assumed [29,33]. Figure 1 summarizes the effective EEA rate constants of monolayer $WS_2$ fabricated on various substrates. The surface roughness of the substrates becomes larger from left to right along the horizontal axis in Fig. 1 (see Supplemental Material). Note that hBN encapsulation is further expected to suppress the intrinsic corrugation [17].

Previously reported data from Refs. 27-29 are also plotted. It should be emphasized that the conventional EEA model is not able to reproduce the time-resolved PL results accurately, and thus the obtained effective EEA rate constants should be considered only as approximate values. Overall, the effective EEA rate constant increases with the surface roughness. This tendency reflects the fact that an increase of the local exciton density due to potential fluctuations enhances the effective EEA rate constant [27]. However, the effective EEA rate constants of the monolayer $WS_2$ on the $SrTiO_3$ and $Al_2O_3$ ST substrates deviate from this tendency and show the lower values. This detail cannot be explained quantitatively with the conventional EEA model. Additionally, such a large difference in the effective EEA rate constant is unattributable to simple environment effects, because dielectric screening from the substrates would only result in an increase of the exciton Bohr radius by about 40% [9,11].

To understand this behavior, we focus on the exciton dynamics in the monolayer $WS_2$ on the $SrTiO_3$ ST substrate, because it had the lowest effective EEA rate constant in the framework of the conventional EEA model. The inset of Fig. 2 shows a typical PL spectrum of this monolayer $WS_2$. The PL peak of the neutral exciton (X) is located at ~2 eV and the tail on the low-energy side is assigned to the negative trion [34]. This assignment is based on the comparison between reflectance contrast and PL spectra (see Supplemental Material). This PL spectrum suggests that the observed PL decay components can be mainly attributed to the exciton emission. Figure 2 shows the normalized PL decay profiles of this sample for different initial exciton densities $N_0$. Here, $N_0$ was determined from the excitation power density per pulse and a linear absorbance of 4%, independent of the substrate [35]. A long decay component (lifetime of about 20 ns) is observed in any case. The observed fast decay component (< 2 ns) is considered to be due to carrier trapping by shallow defect states [28,36] or relaxation to the dark exciton state located ~30 meV below the bright exciton state [37,38]. For low initial exciton densities, the decay is governed by the long decay component. With an increase in excitation laser fluence, a fast-decaying component appears. As will be discussed below, we revealed that the PL components with the quantized lifetimes appear sequentially as increasing in initial exciton density. Our analysis of the fluence dependence of the PL dynamics with the conventional EEA model (see Supplemental Material) shows that the PL decay dynamics under strong and weak excitation conditions cannot be reproduced by the same EEA rate constant. This supports that another model is required to explain our experimental data.

In the conventional EEA model, a spatially homogeneous exciton density is assumed. However, monolayer materials are sensitive to the surrounding environment and thus a spatial inhomogeneity of the electronic potential can occur due to various factors. Although an ST substrate with an atomically flat surface rarely causes a random fluctuation of the electronic potential (which would induce a local accumulation of excitons), a weak and periodic modulation of the potential would be induced by the intrinsic corrugation of the monolayer [39,40] or the one-unit cell step of the substrate. For simplicity, we assume that the monolayer $WS_2$ on the $SrTiO_3$ ST substrate is divided into areas with equal size by the potential barriers induced by the periodic potential modulation. Within each area, the potential is assumed to be constant. In this situation, multi-exciton relaxation dynamics are quantized in terms of the number of excitons $n$ existing in a particular area with size $a$ at a certain time $t$. Here we refer to this model as quantized EEA (QEEA) model. A similar quantization of recombination dynamics has been reported for semiconductor quantum dots [41], quantum rods [42], and carbon nanotubes [43–45]. When the number of excitons is small, the relaxation process is described by a sequence of quantized steps from $n$ to $n−1$, $n−2$, … until zero excitons exist in this particular area [Fig. 3(a)]. In the experiments, we observe the ensemble average of the exciton recombination dynamics determined by the initial number of excitons in each area [Fig. 3(b)].

Two types of exciton decay processes have to be considered: the linear exciton recombination without interactions with other excitons and the nonlinear EEA process involving multiple excitons,

$$E_n \xrightarrow{kn} E_{n-1}, \text{(Linear)}$$

$$E_n \xrightarrow{\frac{n(n-1)}{2}\gamma_{QEEA}} E_{n-1}, \text{(Nonlinear)}$$

where $k$ and $\gamma_{QEEA}$ are the linear recombination rate and the QEEA rate, respectively. $E_n$ refers to a state of an area with $n$ excitons. Note that an exciton that has received additional energy from another exciton in the EEA process, quickly relaxes to the lowest exciton state by emitting phonons. The time needed for such a thermalization process is typically on the order of picoseconds [46,47], which is much shorter than the EEA time constants and exciton recombination lifetimes in this study. The time-dependent probability density $\rho_n$ of finding $n$ excitons in a particular area at time $t$ obeys the following equation [48]:

$$\frac{d}{dt}\rho_n(t) = -\left[k + \frac{1}{2}(n-1)\gamma_{QEEA}\right]n\rho_n(t) + \left(k + \frac{1}{2}n\gamma_{QEEA}\right)(n+1)\rho_{n+1}(t).$$

The average number of excitons per area is described by

$$\bar{n}(t) = \sum_{n=1}^{\infty} n\rho_n(t).$$

The above two equations can be solved by using the generating function technique [48,49]. The solution is given by

$$\bar{n}(t) = \sum_{n=1}^{\infty} \exp\left[-\frac{1}{2}\gamma_{QEEA}n(n+z-1)t\right](z+2n-1)\bar{n}_0^n \sum_{j=0}^{\infty} P_j(\bar{n}_0)\frac{\Gamma(z+n+j)}{\Gamma(z+2n+j)},$$

where $z = 2k/\gamma_{QEEA}$, $P_j(\bar{n}_0)$ is the Poisson distribution function, and the average number of initial excitons is defined by $\bar{n}_0 = \bar{n}(0) = N_0 a$. Note that the decay of the PL intensity for an average number of initial excitons $\bar{n}_0$, $I(\bar{n}_0, t)$, is proportional to $\bar{n}(t)$, and can be written as

$$I(\bar{n}_0, t) = \sum_{n=1} A_n(\bar{n}_0) \exp\left(-\frac{t}{\tau_n}\right).$$

Here, $A_n(\bar{n}_0)$ and $\tau_n$ correspond to the ensemble signal intensity and relaxation lifetime of a decay curve of an $n$-exciton state, respectively. To evaluate the experimental results with the above equation, we used a simple subtractive procedure similar to that used for semiconductor quantum dots [41]. We extracted the decay components up to the three-exciton state and evaluated the $\bar{n}_0$ dependence of $A_n$ for each decay component (see Supplemental Material).

Figure 3(c) shows the decay curves of the one-, two-, and three-exciton states derived by the subtractive procedure. These decay curves were fitted to a single exponential function and as a result we obtained $\tau_1 = 23.3$ ns, $\tau_2 = 7.3$ ns, and $\tau_3 = 3.0$ ns. Within the QEEA model, the inverse lifetime of an $n$-exciton state is expressed by

$$1/\tau_n = kn + {}_nC_2 \gamma_{QEEA}.$$

Therefore, we can determine the $k$ and $\gamma_{QEEA}$ by fitting the above three data points to Eq. (7): $k = 0.042 \pm 0.007$ and $\gamma_{QEEA} = 0.055 \pm 0.018$ ns$^{-1}$. The inset of Fig. 3(c) compares the inverse lifetimes of the $n$-exciton states that were obtained directly with the single exponential functions and those re-calculated using the fitting results of $k$ and $\gamma_{QEEA}$. The good agreement between these values represents the quality of the fitting with Eq. (7) and strongly supports the validity of the QEEA model.

The fitting parameters $A_1$, $A_2$, and $A_3$ are summarized in Fig. 3(d) as a function of $\bar{n}_0$. The broken lines are theoretical predictions from the QEEA model, which reproduce the experimental results well. This result shows that the QEEA model can explain the exciton dynamics in the monolayer WS$_2$ on the SrTiO$_3$ ST substrate. Note that each theoretical curve is multiplied by a constant value to fit the experimental data (see Supplemental Material). The multiplication factor is needed due to the existence of the fast trapping process. By using the QEEA model, we can estimate the average size of individual areas separated by potential barriers. From the effective initial exciton density that takes the fast trapping into account (see Supplemental Material), we can estimate that the average size of the area defined by the potential barriers is about $1.5 \times 10^{-10}$ cm². This corresponds to a characteristic length of $l \approx 100$ nm, which is larger than the mean peak separation of the surface roughness of the commonly used substrates and is about the same size as (or slightly smaller than) the terrace width of the ST substrates. Therefore, we consider that the periodic potential modulation does not originate from the random roughness of the substrate, but from the intrinsic or artificial corrugation of the monolayer $WS_2$ and the periodic structure formed by the steps and terraces of the ST substrate.

The intrinsic EEA rate constant $\gamma$, which corresponds to the value in the conventional EEA model when the electronic potential is homogeneous, is calculated as follows:

$$\gamma = \gamma_{QEEA} a = (8.3 \pm 2.7) \times 10^{-3} \text{ cm}^2 \text{ s}^{-1}.$$

Interestingly, this value is in good agreement with the value of $(6.3 \pm 1.7) \times 10^{-3}$ cm² s⁻¹, which is the previously reported EEA rate constant of hBN-encapsulated monolayer $WS_2$ [27]. In addition, this value is one order of magnitude higher than the effective EEA rate constant obtained from the conventional EEA model. We were able to estimate the intrinsic EEA rate constant by using the QEEA model and considering periodic potential fluctuations. This analysis clarifies that the unexpected difference between the values obtained for hBN-encapsulation and the $SrTiO_3$ ST substrate in Fig. 1 was caused by using a model that contains assumptions that are inappropriate for ST substrates. It should be noted that the QEEA model can be also applied to the results for the $Al_2O_3$ ST substrate, which further supports the validity of this model. On the other hand, the QEEA model cannot account for the EEA processes in monolayer $WS_2$ on substrates with strong random roughness, such as the $SiO_2$/Si substrate (see Supplemental Material). A large potential fluctuation induced by random roughness can cause a local increase in the exciton density, which enhances the effective EEA rate constant as reported previously [27,28] and confirmed in Fig. 1. Therefore, we consider that the effective EEA rate constant is determined by two competing factors: the QEEA process and the local increase in the exciton density. Based on this idea, we can qualitatively explain the substrate dependence of the effective EEA rate constant shown in Fig. 1. Further research is needed to obtain a unified model in which different mechanisms act according to the shape of the potential fluctuation.

In conclusion, we have discussed the influences of potential fluctuations on the nonlinear PL dynamics dominated by EEA processes in monolayer $WS_2$ on various substrates. We have shown that the effective EEA rate constant significantly depends on the potential fluctuation caused by the roughness of the substrate surface. In particular, the effective EEA rate constant that is slightly lower than that of hBN-encapsulated $WS_2$ was found for the monolayer $WS_2$ on the $SrTiO_3$ substrate with atomically flat terraces. We proposed the QEEA model to explain the PL decay curves with discrete lifetimes as a result of excitons being confined to individual areas by potential barriers. Here, the barriers were considered to be due to intrinsic or artificial corrugation. The QEEA model well reproduced the relaxation lifetimes up to the three-exciton state and also the dependence of the PL intensity on the average number of the initial excitons. Furthermore, the intrinsic EEA rate constant of the monolayer $WS_2$ was estimated with this model. The present results highlight the importance of the potential fluctuations in monolayer materials for multi-exciton dynamics and may help to develop a method to suppress the EEA process by using spatial modulation in heterostructures, such as Moiré and superlattice structures.

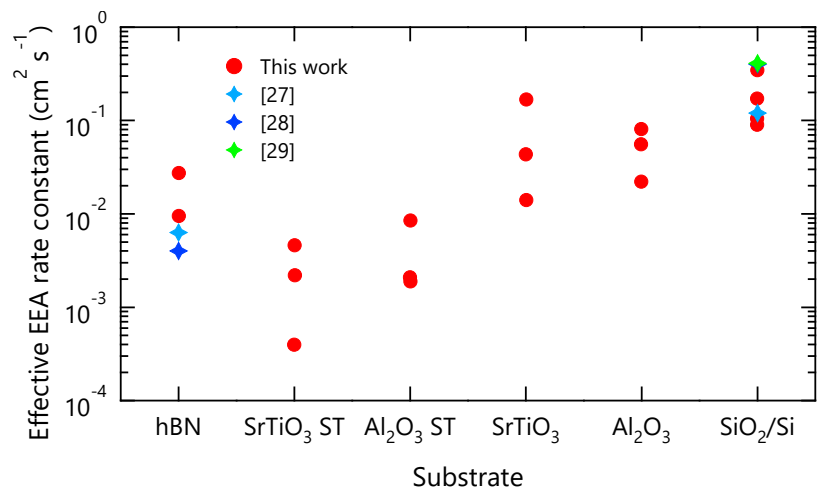

FIG. 1. (Color Online)

Effective EEA rate constants of monolayer $WS_2$ on various substrates.

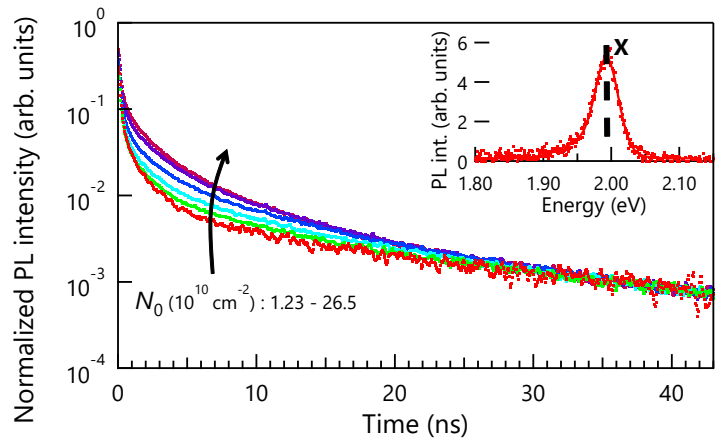

FIG. 2. (Color Online)

PL decay profiles of the monolayer $WS_2$ on the $SrTiO_3$ ST substrate. Data for six different initial exciton densities are provided. The decay curves are normalized by their intensities at around 40 ns. The inset shows a typical PL spectrum and the dashed line indicates the exciton PL peak energy.

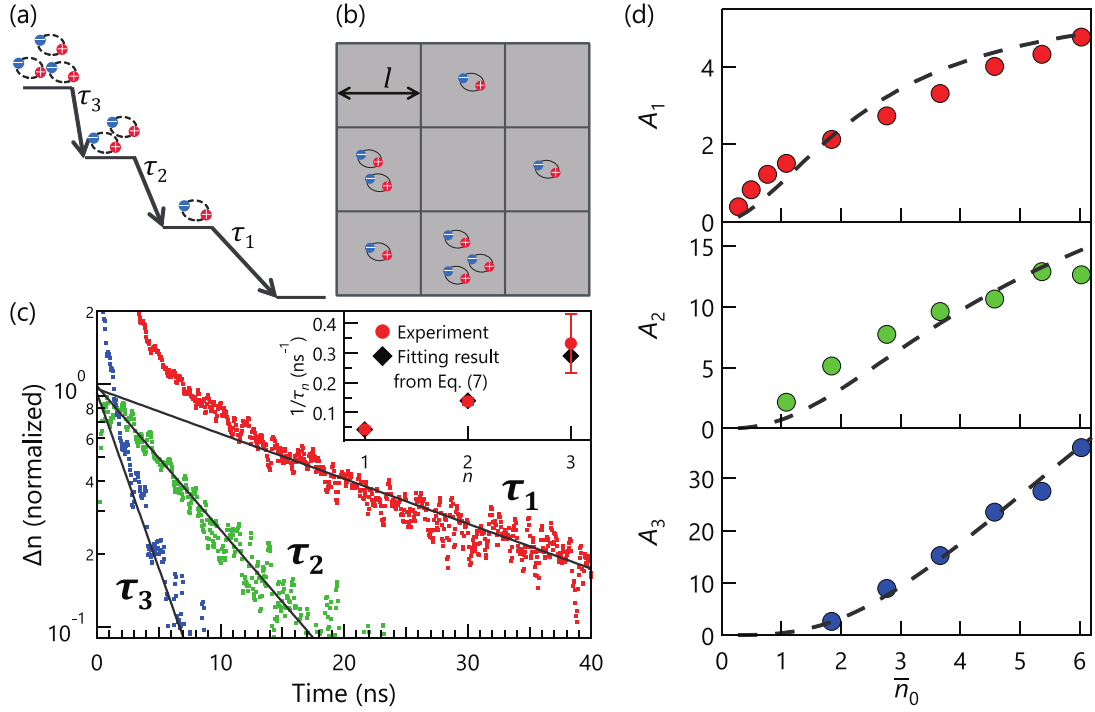

FIG. 3. (Color Online)

(a) Schematic of the quantized relaxation dynamics starting from a three-exciton state in the QEEA model. (b) Illustration of monolayer $WS_2$ divided into areas with equal sizes due to potential barriers, and different numbers of excitons in each area. (c) Decay curves of the one-, two-, and three-exciton states extracted using a simple subtractive procedure. The solid lines are the fitting results obtained with a single exponential function. The inset shows the inverse of the obtained lifetimes of the $n$-exciton states. (d) Fitting parameters $A_1$, $A_2$, and $A_3$ as a function of the average number of initial excitons. The broken curves are the theoretical curves predicted by the QEEA model.